\documentclass[letterpaper, 10 pt, conference]{ieeeconf}  

\IEEEoverridecommandlockouts                              
\usepackage{graphicx} 
\usepackage{amssymb}
\usepackage{epsfig} 
\usepackage{amssymb}  
\usepackage{titlecaps}
\usepackage{multirow,booktabs} 
\usepackage{changes} 
\setremarkmarkup{(#2)} 
\usepackage{xcolor}
\definecolor{darkgreen}{rgb}{0.0, 0.4, 0.2}
\definechangesauthor[name={Matthew Fricke}, color=darkgreen]{GMF}
\definechangesauthor[name={Melanie Moses}, color=orange]{MEM}
\definechangesauthor[name={Sarah Ackerman}, color=blue]{SMA}

\usepackage{subcaption} 
\usepackage[normalem]{ulem} 
\usepackage{siunitx}
\usepackage[noadjust]{cite}

\usepackage[hyphens]{url}

\usepackage{subcaption}
\usepackage[hang,flushmargin,symbol]{footmisc} 
\usepackage{algorithm,algorithmicx,algpseudocode}
\usepackage{csquotes} 
\usepackage[all]{nowidow}
\usepackage{acro} 

\makeatletter 
\algnewcommand{\LineComment}[1]{\Statex \hskip\ALG@thistlm \(\triangleright\) #1}

\Addlcwords {a an the at by for in of on to up and as but or nor so yet vs with per} 

\DeclareAcronym{ISRU}{
  short = ISRU,
  long = in-situ resource utilization,
}

\DeclareAcronym{NASA}{
  short = NASA,
  long = National Aeronautics and Space Administration,
}

\DeclareAcronym{ROS}{
  short = ROS,
  long = the robot operating system,
  cite = {Quigley2009}
}

\DeclareAcronym{CS}{
  short = CS,
  long = computer science,
}
\DeclareAcronym{GPS}{
  short = GPS,
  long = global positioning system,
}

\DeclareAcronym{IMU}{
  short = IMU,
  long = inertial measurement unit,
}

\DeclareAcronym{PID}{
  short = PID,
  long = proportional-integral-derivative,
}

\DeclareAcronym{MSI}{
  short = MSI,
  long = minority-serving institution,
}

\DeclareAcronym{CPF}{
  short = CPF,
  long = central-place foraging,
}

\DeclareAcronym{MUREP}{
  short = MUREP,
  long = Minority University and Research Education Program,
}

\DeclareAcronym{STEM}{
  short = STEM,
  long = {science, technology, engineering, and mathematics},  
}

\DeclareAcronym{ABET}{
  short = ABET,
  long = Accreditation Board for Engineering and Technology,
}

\DeclareAcronym{EKF}{
  short = EKF,
  long = extended Kalman filter,
  cite={Moore2014}
}

\DeclareAcronym{GUI}{
  short = GUI,
  long = graphical user interface,
}

\DeclareAcronym{SIPI}{
  short = SIPI,
  long = Southwestern Indian Polytechnic Institute,
}

\DeclareAcronym{UNM}{
  short = UNM,
  long = the University of New Mexico,
}

\makeatother
\title{\LARGE \bf
\titlecap{The Swarmathon: An Autonomous Swarm Robotics Competition}
}

\author{Sarah M. Ackerman$^{*1}$, G. Matthew Fricke$^1$, Joshua P. Hecker$^1$, Kastro M. Hamed$^2$, Samantha R. Fowler$^2$,\\ Antonio D. Griego$^1$, Jarett C. Jones$^1$, J. Jake  Nichol$^{1}$, Kurt W. Leucht$^3$, and Melanie E. Moses$^{1,4,5}$
\thanks{$^1$Computer Science Department, The University of New Mexico, Albuquerque, NM, USA, $^2$Education and Interdisciplinary Studies, Florida Institute of Technology, Melbourne, FL, USA, $^3$NASA Kennedy Space Center, FL, USA, $^4$Department of Biology, The University of New Mexico, $^5$Santa Fe Institute, Santa Fe, NM, USA, This work was supported by NASA \ac{MUREP} \#NNX15AM14A and a generous donation from Google. We thank Theresa Martinez, \ac{MUREP} STEM Engagement Manager; Paul Secor and Mary Baker of Secor Strategies, LLC; Kate Cunningham and Beatriz Palacios Abad of UNM ADVANCE; Elizabeth Esterly and the students of the Moses Biological Computation Lab, Vanessa Svihla, Aeron Haynie, NASA volunteers, UNM support staff, KSC Visitor Complex, and the Swarmathon students and faculty mentors.}
\thanks{$^*$Correspondence: \texttt{sherbet@unm.edu}}
}

\begin{document}

\maketitle
\thispagestyle{empty}
\pagestyle{empty}

\begin{abstract}

The Swarmathon is a swarm robotics programming challenge that engages college students from minority-serving institutions in NASA's \textit{Journey to Mars}. Teams compete by programming a group of robots to search for,  pick up, and drop off resources in a collection zone. The Swarmathon produces prototypes for robot swarms that would collect resources on the surface of Mars. Robots operate completely autonomously with no global map, and each team's algorithm must be sufficiently flexible to effectively find resources from a variety of unknown distributions. The Swarmathon includes Physical and Virtual Competitions. Physical competitors test their algorithms on robots they build at their schools; they then upload their code to run autonomously on identical robots during the three day competition in an outdoor arena at Kennedy Space Center. Virtual competitors complete an identical challenge in simulation. Participants  mentor local teams to compete in a separate High School Division. In the first 2 years, over 1,100 students participated. 63\% of students were from underrepresented ethnic and racial groups. Participants had significant gains in both interest and core robotic competencies that were equivalent across gender and racial groups, suggesting that the Swarmathon is effectively educating a diverse population of future roboticists.

\end{abstract}


\section{INTRODUCTION}

\subsection{\titlecap{The Journey to Mars}}
The \ac{NASA} \textit{Journey to Mars} program has set the ambitious goal of sending manned missions to Mars by the 2030s.\footnote{\url{https://www.nasa.gov/content/nasas-journey-to-mars}} These surface missions may last months or years, and because transporting sufficient materials from Earth is not practical, astronauts will need to utilize the resources already available on Mars. For example, small pockets of ice can be melted for water and converted to oxygen and hydrogen for fuel. This approach is known as \ac{ISRU}. 

\subsection{\titlecap{Meeting the Challenge with Robot Swarms}}

Autonomous robot swarms could locate, collect, and store resources in advance of human arrival. A swarm of small, inexpensive robots acting autonomously provides several advantages over a few large, expensive, manually controlled rovers \cite{Brambilla2013,Zedadra2017} because they are robust, scalable and able to explore unmapped environments \cite{hecker2012formica}.

\ac{ISRU} is closely related to \ac{CPF}, in which resources are transported to a central collection zone. Inspired by the success of social insects gathering resources for their colonies, an early study of robot foraging computationally modeled a swarm collecting rock samples on a distant planet \cite{Steels1990}. \ac{CPF} is a key robot swarm application \cite{Drogoul1993,Campo2007}, and research continues to improve algorithms and engineering approaches for foraging swarms
\cite{Winfield2009a,Dorigo2014}.
  
Our team has designed and built a swarm of foraging robots called \textit{Swarmies}. Swarmies are rugged enough to operate outdoors for hours, and feature sensors and grippers that enable them to complete CPF tasks. We also developed custom software packages and a \ac{GUI}, enabling rapid development and testing of CPF algorithms in simulation and physical hardware.

Swarm foraging provides an ideal educational challenge because successful strategies require foundational robotics skills such as grasping, localization, navigation, exploration, and decentralized communication and coordination \cite{Winfield2009}.

\subsection{The Swarmathon Competition}

In the Swarmathon annual competition, teams of college students develop algorithms for autonomous swarm foraging. Teams in the Physical Competition program groups of 3 - 6 Swarmies to collect the most of $128 - 256$ resources placed in $225 - \SI{484}{\meter\squared}$ outdoor arenas. The Virtual Competition runs in a Gazebo simulation of the same arenas and robots. Robots must search for, collect, and return resources to a central collection zone completely autonomously. The competition runs in a series of rounds, and in each round resources are placed in different distributions (i.e., uniformly at random, or in various sizes and shapes of clusters), and the same code is run in each round. Score is determined by the number of resources in the collection zone at the end of the round (resources pushed out of the collection zone by robots do not count toward the score). Robots have limited view and no global map, making it difficult to find resources. In a third High School Competition, younger students are mentored by Swarmathon college students to complete a similar challenge in a simplified simulation. Winners of the Virtual competition advance to the Physical competition the following year. 

\subsection{\titlecap{ Diversifying the STEM/Robotics Pipeline}}
Tackling \ac{NASA}'s real-world \ac{ISRU} problem provides an opportunity to engage students in robotics and \ac{CS}, helping meet the growing need for \ac{STEM} talent in the workforce. While women and minorities account for a growing proportion of US college graduates and US workers, they are underrepresented among \ac{STEM} college graduates and professionals, especially in \ac{CS} and engineering \cite{Griffith2010,Leggon2006,Museus2011,USDepartmentofEducation2016}. Their inclusion in STEM helps to address an anticipated shortfall in these fields over the next few decades \cite{Museus2011,USDepartmentofEducation2016}. Additionally \enquote{workers from a variety of backgrounds enhance the quality of science insofar as they are likely to bring a variety of new perspectives\ldots in terms of both research and application}  \cite{Leggon2006}. For these reasons, the Swarmathon recruits competitors from 2- and 4-year colleges and universities that are \acp{MSI} including  Historically Black Colleges and Universities (HBCUs), Hispanic-Serving Institutions (HSIs), and Tribal Colleges and Universities (TCUs). This provides an educational opportunity that may not otherwise exist for participating students:

\blockquote{``The Swarmathon is an opportunity for students to work on a real-life engineering problem that's interdisciplinary and hard. They know that I don't know the answer which makes it fun for all of us. I teach at a small community college and it simply wouldn't be possible for us do this level of work from scratch.'' -Swarmathon Faculty Mentor}

The Swarmathon has engaged a diverse student population with 81\% of students from minority groups (including Asian students) and 63\% of the 818 college participants identifying as belonging to an underrepresented racial or ethnic group (Black/African American, American Indian/Alaskan Native, or Hispanic/Latino), substantially more diverse than US undergraduate \ac{CS} majors, of which only 15.8\% are from these underrepresented groups \cite{Zweben2017}. 

College students mentored K-12 students in their communities, including 300 high school students who compteed in a simplified High School Swarmathon in Netlogo. Additionally, the Swarmathon supported 60 students to attend workshops and the 2016 and 2017 Robotics Science and Systems (RSS) conferences and 17 students participated in research internships with mentors as other Swarmathon schools. All of these activities further developed pathways into STEM. As an example, 65\% of RSS workshop participants indicated a desire to pursue a Ph.D. in robotics. 

\section{RELATED WORK}

\subsection{\titlecap{Foraging Robot Swarms}}

Though swarm robot foraging has been studied analytically for decades \cite{Steels1990}, hardware implementations are rare \cite{Winfield2009}. For example, studies of \ac{CPF} often only simulate the collection and transport of objects \cite{Brutschy2015}, and recent advances in hardware implementations, such as collaborative warehouse robots, require permanent infrastructure such as buried guide-wires or visual markers to operate \cite{Enright2011}. 

Many attempts have been made to develop swarm algorithms and robot systems that address various aspects of the swarm foraging problem including scalability, energy efficiency, task allocation, and collection speed \cite{Liu2006,hoff2010two,Hecker2015,Ferrante2015,Lu2016,Fricke2016c}. However, autonomous swarm foraging remains a challenging open problem with no generally recognized best solution.

Projects using physical robot swarms include the Robotarium, a swarm robotics testbed providing remotely accessible robots and an arena \cite{Pickem2017}. This innovative project allows virtual experimentation with physical robot swarms, but it uses an overhead camera for localization and the robots themselves have no on-board sensors. Kilobots are simple, small robots intended to be integrated into large swarms. They have sensors and operate autonomously and collaboratively to push items through a maze or into specific configurations, but they have relatively limited mobility and operate in controlled laboratory environments \cite{Rubenstein2014}. The Swarmanoid project demonstrated successful implementation of a heterogeneous swarm whose robots collaborate in order to solve tasks like object retrieval in a highly specialized environment \cite{Dorigo2013}.
  The challenges of operating physical swarms result in robots that are usually {\itshape semi-autonomous} in practice; they require frequent human support \cite{Rosenfeld2017}. The \textit{reality gap} between performance in simulation and performance in physical hardware \cite{Jakobi1995} makes implementation in real robots particularly difficult. The hardware and software platforms for the Swarmies address these challenges so that \ac{CPF} can be developed and subsequently tested in a system that is completely autonomous, operates without global knowledge or control, and can function outdoors.

\subsection{\titlecap{Educational Robotics Environments and Competitions}}
Educational robotics is a growing field that is particularly effective at improving student performance in and attitudes towards STEM disciplines, and reducing gender and ethnic achievement gaps 
\cite{Freeman2014,Museus2011, Welch2010,Richmond2016,Steele2010}. Examples for younger students include the distributed Robot Garden \cite{Sanneman2015} and low-cost AERobots \cite{Rubenstein2015}. 

The \ac{NASA} Robotic Mining Competition (RMC) is an annual engineering competition to design and build robots capable of mining the Martian surface as part of the \ac{ISRU} mission. Students focus on hardware design rather than autonomous control. Only 16\% of the 900 RMC participants in 2017 identified as members of underrepresented groups\footnote{\url{https://www.nasa.gov/offices/education/centers/kennedy/technology/nasarmc.html}}. The Swarmathon was designed to emulate the success and popularity of RMC while emphasizing robot autonomy and engagement of students from \acp{MSI}.

\subsection{\titlecap{Hardware}}

Teams selected for the Physical Competition are sent parts and instructions\footnote{\url{https://github.com/BCLab-UNM/Swarmathon-Robot}} for building 3 Swarmies.  
Each robot has a front mounted gripper with actuated wrist and finger joints for grasping resources. Robots sense objects in the environment using three ultrasound range finders and a web camera. The ultrasounds have a \SI{3}{\meter} range. The camera has a narrow field of view with a \SI{1}{\radian} arc and range of \SI{1}{\meter}.

Orientation and positional (pose) data are provided by a \ac{GPS} unit and \ac{IMU} with a magnetometer, along with wheel encoder odometry. Computation is provided by a small on-board computer (an Intel NUC). An Arduino Leonardo microcontroller provides the hardware interface between the on-board computer and both the wheel and gripper actuators and the  encoders that measure wheel velocity. The robot body is built from laser cut and 3D printed components. The battery pack allows robots to run for 8 hours between charges, which with a default speed of speed of \SI{0.2}{\meter/\second} results in a range of \SI{5.75}{\kilo\meter}. This allows a team of 6 Swarmies to search a linear distance of \SI{34.5}{\kilo\meter} on a single charge, nearly the distance of a marathon that the remote-controlled Mars Rover Opportunity took 11 years to complete.

The resources that Swarmies locate and collect are cubes marked on all sides with AprilTags \cite{Olson2011}. The perimeter of the central collection zone is also marked with AprilTags. 

\subsection{\titlecap{Software}}

    \begin{figure}
      \centering
      \includegraphics[width=1.0\columnwidth]{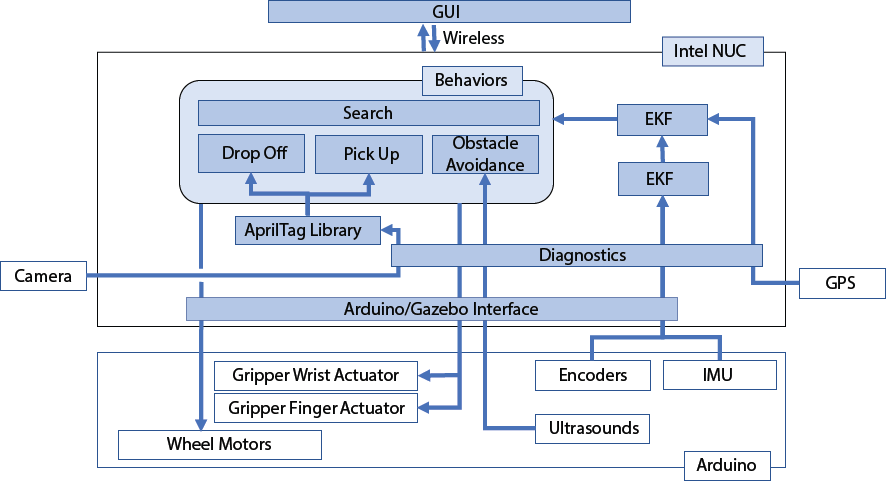}
      \caption{Swarmie Architecture. Software packages (shaded) interface  with hardware through a USB connection, or with the Gazebo simulation.}
      \label{fig:swarmie_arch}
   \end{figure} 

The basic robot software, based on \ac{ROS}, is publicly available\footnote{\url{https://github.com/BCLab-UNM/Swarmathon-ROS}} and provided to all teams. So teams may focus on the \ac{ISRU} challenge, we implement the low-level code that interfaces with physical and simulated actuators and sensors, performs necessary calibration and pre-configuration, and defines the simulated world for competition, including Swarmies, resources, and the search environment. 

The software base is designed so that students can implement the same algorithms to control simulated or physical robots (Figure \ref{fig:swarmie_arch}). The \ac{GUI} and \ac{ROS} master either connect to the physical robots through a wireless network, or run in a Gazebo simulation. The diagnostic package monitors hardware components and alerts the user to problems. The robot pose is estimated using two \acp{EKF} that fuse encoder, \ac{IMU}, and \ac{GPS} data.

\begin{figure*}[ht]
	\begin{subfigure}{0.275\textwidth}
      \centering
      \includegraphics[width=\textwidth]{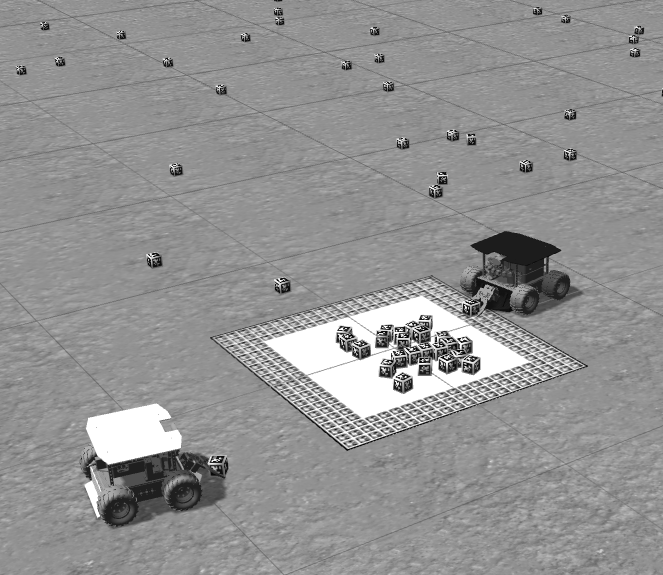}
      \caption{Gazebo Simulation}
      \label{fig:sim_robot}
      \end{subfigure}
      ~ 
      	\begin{subfigure}{0.4\textwidth}
      \centering
      \includegraphics[width=\textwidth]{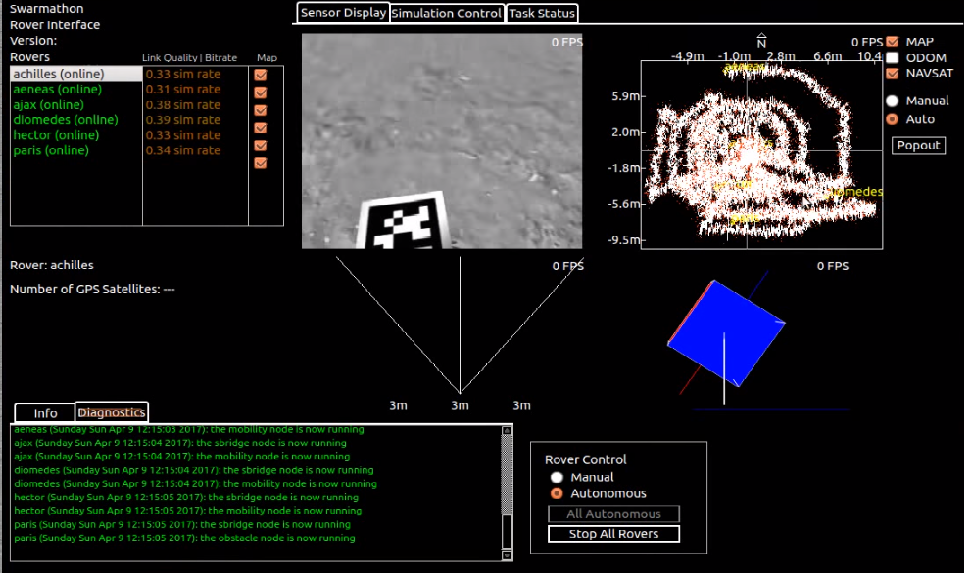}
      \caption{User Interface}
      \label{fig:GUI}
      \end{subfigure}
      ~ 
      	\begin{subfigure}{0.3\textwidth}
      \centering
      \includegraphics[width=\textwidth, height=4cm]{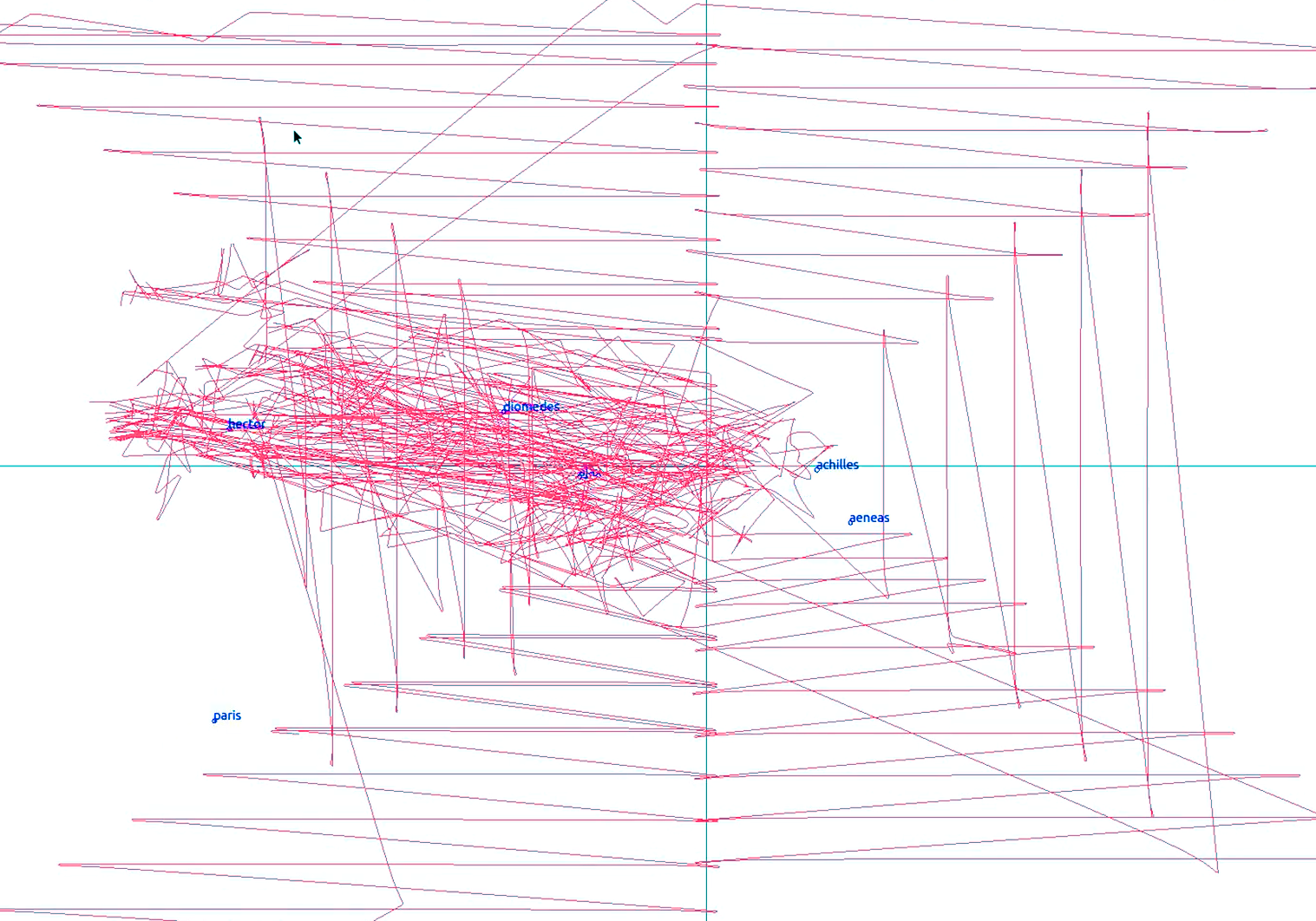}
      \caption{Example Search Pattern}
      \label{fig:example_search_pattern}
      \end{subfigure}
      \caption{Gazebo and GUI. (a) Two robots bring target cubes marked with AprilTags to the collection zone.  (b) Gazebo interface showing robot status, location tracing, and camera view. (c) Trace of the Montgomery College winning swarm of 6 robots; the dense activity shows robots traveling between a large target cluster and the collection zone.}
      \label{fig:CombinedFigure}
   \end{figure*}

This base code includes: Gazebo simulation of the robots, target cubes, and competition arena (Figure \ref{fig:CombinedFigure}); a \ac{GUI} (Figure \ref{fig:GUI}) showing output of each hardware sensor, a map of the robots' estimated positions (Figure \ref{fig:example_search_pattern}); customizable ROS packages to control behaviors (for search, communication, obstacle avoidance, and target pick up and drop off) and diagnostics (for sensors, actuators, wireless quality and the microcontroller); a \ac{ROS} package that interfaces with the Arduino to communicate with actuators and  and sensors; open source packages including AprilTag and \ac{EKF} packages.

The simple base code algorithm (modeled after the subsumption architecture \cite{Brooks1986}) is designed to be extended by teams to improve foraging. It implements simple collision avoidance, a random walk for search, and when a target appears in the camera view, the robot picks up the target and returns to the central collection zone to drop it off.

\begin{figure}
      \centering
      \includegraphics[width=1.0\columnwidth]{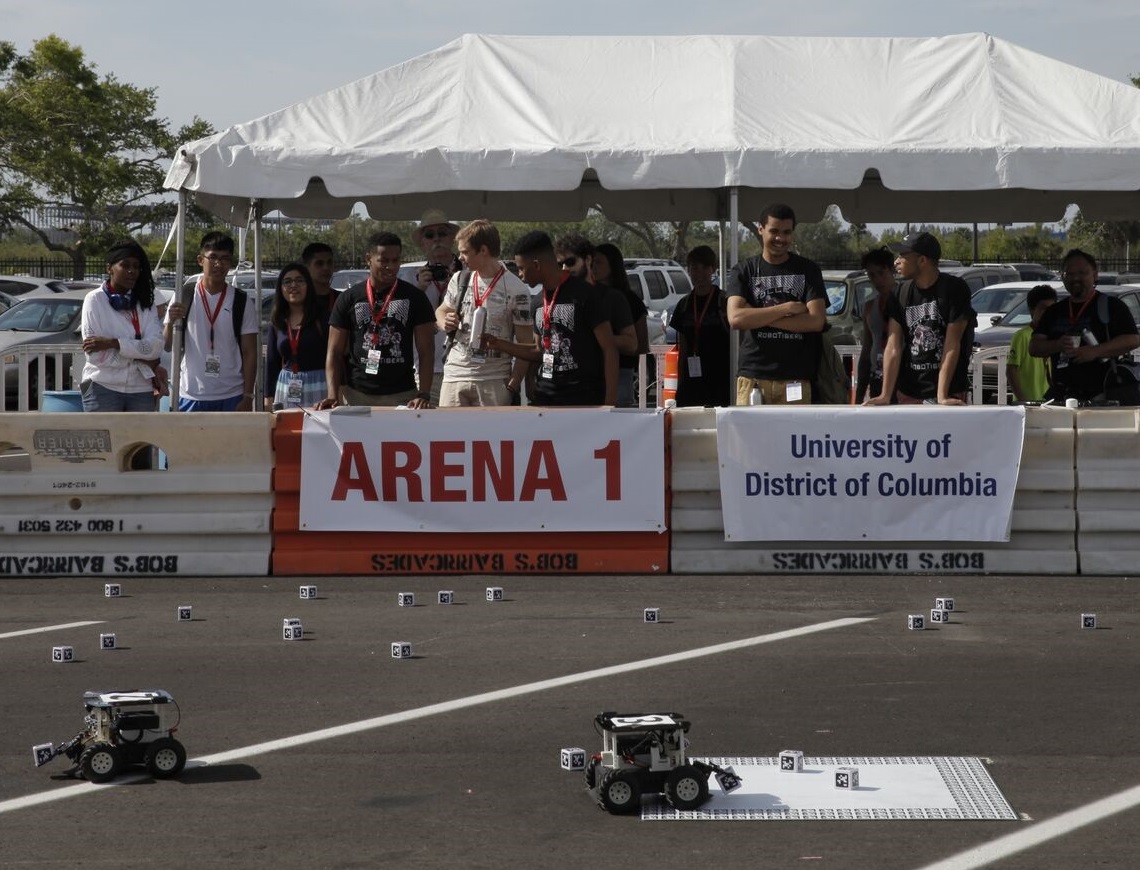}
      \caption{Physical Competition. Students compare strategies and cheer on Swarmies from the University of the District of Columbia as they drop cubes in the collection zone.}
      \label{fig:arenapic}
   \end{figure}

\section{RESULTS}

In the first 2 years, 29 of the 31 Physical teams and 20 of the 27 Virtual Teams successfully uploaded code to compete. The skills these students required for the competition included programming proficiency in \ac{ROS}, swarm algorithm design, and experimental design and testing. 

Several teams developed effective new search algorithms that significantly improved the base code. The 2017 Physical Competition winner, \ac{SIPI}, experimented with different search strategies inspired by snail shells, lawnmower paths, and spirograph-style geometric designs, ultimately deciding on a strategy inspired by the spokes of a bicycle wheel. \ac{SIPI}'s algorithm displayed the greatest flexibility, performing at a high level across a variety of target distributions.

Figure \ref{fig:example_search_pattern} shows the Montgomery College search algorithm in an environment with one large cluster of target cubes. The zig-zag pattern shows the robots' search path through the arena. Once the large cluster of resources was located, some robots continued searching while others focused on collecting from the cluster. This strategy successfully balanced the explore versus exploit trade-off and dramatically outperformed the other teams on nearly every distribution of resources to win the 2017 Virtual Competition. Engineering innovations (i.e., improving localization and resource pickup and drop-off) improved scores more than new search algorithms.

\subsection{\titlecap{Assessed Student Learning Outcomes}}

263 students completed a comprehensive self-assessment (29\%). 44 Faculty Mentors (76\%) completed the survey with the same questions to evaluate their students' performance. 

Student growth was rated on 9 \ac{ABET} Criteria for Accrediting Computing Programs, and 2 \ac{ABET} Criteria for Accrediting Engineering Programs: \enquote{An ability to design and conduct experiments, as well as to analyze and interpret data,} and, \enquote{An ability to function on multidisciplinary teams.}\footnote{\url{www.abet.org/accreditation/accreditation-criteria/}} Students and faculty rated student degree of competency before and after the Swarmathon on a 4-point scale: none (1), low degree (2), moderate degree (3), and high degree (4).

Data for each year was analyzed with a dependent-samples t-test. Differences between the 2016 and 2017 cohorts were analyzed with an independent samples t-test. 
Table \ref{tab:surveygains} shows the significant gains in self-assessed performance each year. 2017 gains were greater than 2016 gains in all categories and all results are statistically significant with $p < 0.05$. 

There were no statistically significant differences in our assessment or survey data based on race, ethnicity, or gender, indicating that the Swarmathon provided a level playing field for competition, eliminating the achievement gap noted in other studies \cite{Steele2010,Richmond2016}. Faculty mentor surveys closely mirrored student self-assessments on all but two ABET standards. Mentors ratings were more than 0.50 points lower than student ratings on (1) An understanding of professional, ethical, legal, security and social issues and responsibilities, and (2) An ability to analyze the local and global impact of computing on individuals, organizations, and society.


\begin{table}[ht]
\caption{\titlecap{Changes in student abilities and attitudes}}
\begin{center}
\label{tab:surveygains}
\resizebox{\columnwidth}{!}{ 
    \begin{tabular}{c c c}
    \toprule
                            & 2016    &    2017 \\ 
    \midrule
    ABET Outcomes           & $+0.35$ & $+0.59$ \\ 
    Higher Degree Interest  & $+0.21$ & $+0.53$ \\ 
    Career Motivation       & $+0.19$ & $+0.38$ \\ 
    \bottomrule
    \end{tabular}
    }
\end{center}
\end{table}

Across both competition years, students also reported significant gains in their \enquote{desire or motivation to pursue a higher degree beyond the one I am currently pursuing,} and their \enquote{desire or motivation to pursue a career in \ac{CS}, or Robotics, or other technical discipline in STEM} as seen in Table \ref{tab:surveygains}. 
Gains were significant even among students who already possessed a moderate to high degree of interest in pursuing a more advanced degree.

The Swarmathon is effectively inspiring the next generation of engineers to tackle major scientific problems like the \ac{NASA} \ac{ISRU} mission and to bring their diversity of background and experience into the \ac{STEM} workforce.

\section*{ACKNOWLEDGMENT}

We thank Theresa Martinez, \ac{MUREP} \ac{STEM} Engagement Manager; Paul Secor and Mary Baker of Secor Strategies, LLC; Kate Cunningham and Beatriz Palacios Abad of UNM ADVANCE; Elizabeth Esterly and the students of the Moses Biological Computation Lab, Vanessa Svihla, Aeron Haynie, the NASA event volunteers and UNM support staff, KSC Visitor Complex, and the Swarmathon students and Faculty Mentors.

\addtolength{\textheight}{-0cm}   








\Urlmuskip=0mu plus 1mu
\bibliography{Mendeley}
\bibliographystyle{IEEEtran}

\end{document}